# An Implicit Cover Problem in Wild Population Study[*]

Mary V. Ashley[†]   Tanya Y. Berger-Wolf[‡]

Wanpracha Chaovalitwongse[§]   Bhaskar DasGupta[‡][¶]

Ashfaq Khokhar[‡]   Saad Sheikh[‡]

November 20, 2018

**Abstract**

In an implicit combinatorial optimization problem, the constraints are not enumerated explicitly but rather stated implicitly through equations, other constraints or auxiliary algorithms. An important subclass of such problems is the implicit set cover (or, equivalently, hitting set) problem in which the sets are not given explicitly but rather defined implicitly For example, the well-known minimum feedback arc set problem is such a problem. In this paper, we consider such a cover problem that arises in the study of wild populations in biology in which the sets are defined implicitly via the Mendelian constraints and prove approximability results for this problem.

## 1 Introduction

In an *implicit* combinatorial optimization problem, the constraints are not enumerated explicitly but rather stated implicitly through equations, other constraints or auxiliary algorithms. Well-known examples of such optimization problems include convex optimization problems where the constraints are not given explicitly but rather can be queried implicitly through a separation oracle or given by an auxiliary algorithm. For example, the ellipsoid method can be used to solve in polynomial time a linear programming problem with possibly exponentially many constraints provided we have a separation oracle that, given a tentative solution, in polynomial time either verifies that the solution is a feasible solution or provides a hyperplane separating the solution point from the feasible region. This paper concerns the implicit set cover problems which are defined as follows. In the standard (unweighted) version of the set cover problem, we are given a collection of subsets $\mathcal{S}$ over an universe of elements $\mathcal{U}$ and the goal is to find a sub-collection of sets from $\mathcal{S}$ of minimum cardinality such that the union of these sets is precisely $\mathcal{U}$. A combinatorially equivalent version of the set cover problem is the so-called hitting set problem where one needs to pick instead a subset of the universe $\mathcal{U}$ of minimum cardinality which contains at least one element from every set. Set cover and hitting set problems are fundamental problems in combinatorial optimization whose computational complexities have been throughly investigated and well understood [11]. More general version of the problem could include generalizing the objective function to be minimized, namely the number of sets picked, by say having weighted sets and minimizing the sum of weights of the selected sets, or by defining a monotone objective function on the set system.

[*]A preliminary version of these results appeared in the 5th International Conference on Algorithmic Aspects in Information and Management, A. Goldberg and Y. Zhou (Eds.), LNCS 5564, 43-54, June 15-17, 2009.

[†]Department of Biological Sciences University of Illinois at Chicago, 840 West Taylor Street, Chicago, IL 60607. Email: `ashley@eeb.uic.edu`.

[‡]Department of Computer Science, University of Illinois at Chicago, 851 South Morgan Street, Chicago, IL 60607. Email: `{dasgupta,ssheikh,tanyabw,ashfaq}@cs.uic.edu`. Supported by NSF grant IIS-0612044.

[§]Department of Industrial Engineering, Rutgers University, PO Box 909, Piscataway, NJ 08855. Email: `wchaoval@rci.rutgers.edu`.

[¶]Supported by NSF grants DBI-0543365 and IIS-0346973.

Implicit set cover (or hitting set) problems have the same standard setting, but the sets are not given explicitly but rather implicitly through some implicit combinatorial constraints. For example, the minimum feedback vertex set or the minimum feedback arc set problems are examples of such implicit hitting set problems. Such implicit set cover or hitting set problems can be characterized by not giving the collection of sets $\mathcal{S}$ explicitly but via an efficient (polynomial-time) *oracle* $\mathcal{O}$ that will supply members of $\mathcal{S}$ satisfying certain conditions. For example, the recent work of Richard Karp and Erick Moreno Centeno[1] considers some implicit hitting set problems with applications to multiple genome alignments in computational biology in which the oracle $\mathcal{O}$ provides a minimum-cardinality set (or a good approximation to it) from the collection $\mathcal{S}$ that is disjoint from a given set $Q$. In addition to standard polynomial-time approximation guarantees, one could also invoke other measures of efficiencies, such as number of access to the oracle $\mathcal{O}$ to obtain an optimal or near-optimal solution to the hitting set problem as used by Karp and Centeno.

In this paper, we consider an implicit (unweighted) set cover problem, which we call the MIN-PARENT problem, that arises in the study of wild population. Our problem in the setting described above is roughly as follows. Our oracle $\mathcal{O}$ returns, given a sub-collection of elements $\mathcal{U}' \subseteq \mathcal{U}$, if there is a set that includes $\mathcal{U}'$. Our specific objective function is motivated by the biological application and is a monotone function, namely including a new element in our collection does not decrease it. More precise formulations of our problems appear in the next section and will easily convince the reader that our problem is not captured by previous works or the recent work by Karp and Centeno.

---

[1] Richard Karp, UC Berkeley, personal communication.

## 2 Motivations

For wild populations, the growing development and application of molecular markers provides new possibilities for the investigation of many fundamental biological phenomena, including mating systems, selection and adaptation, kin selection, and dispersal patterns. In our motivation we are concerned with full sibling relationships from single generation sample of microsatellite markers. Several methods for sibling reconstruction from microsatellite data have been proposed (*e.g.*, see [6, 10, 12]). Combinatorial approaches to sibling reconstruction using suitable parsimony assumptions have been studied in [2–5]. These approaches use the Mendelian inheritance rules to impose constraints on the genetic content possibilities of a sibling group. A formulation of the inferred combinatorial constraints in constructing a collection of groups of individuals that satisfy these constraints under the *parsimony* assumption of *a minimum number of parents* leads to the MIN-PARENT problems discussed in the paper.

## 3 Precise Formulations of MIN-PARENT Problems

An element (*individual*) $u$ is an ordered sequence $(u_1, u_2, \ldots, u_\ell)$ where each $u_j$ is a genetic trait (*locus*) and is represented by a multi-set $\{u_{j,0}, u_{j,1}\}$ of two (possibly equal) numbers (*alleles*) inherited from its parents. Biologically, each element corresponds to an individual in the sample of the wild population from the *same* generation. We have a universe $\mathcal{U}$ consisting of $n$ such elements. Certain sets of individuals in $\mathcal{U}$ can be *full siblings*, *i.e.* having the *same* pair of parents under the *Mendelian inheritance rule. These sets are specified in an implicit manner in the following way.* The Mendelian inheritance rule states that an individual $u = (u_1, u_2, \ldots, u_\ell) \in \mathcal{U}$ can be a child of a pair of individuals (*parents*), say $v = (v_1, v_2, \ldots, v_\ell)$ and $w = (w_1, w_2, \ldots, w_\ell)$, if and only if for each locus $j \in \{1, \ldots, \ell\}$ one allele of $u_j$ is from $v_j$ and the other element of $u_j$ is from $w_j$. Finally, a subset $\mathcal{U}' \subseteq \mathcal{U}$ is a (*full*) *sibling group* if and only if there exists a pair of parents $v$ and $w$ such that every member of $\mathcal{U}'$ is a child of $v$ and $w$. Note that *any* pair of individuals is a full sibling group by the Mendelian constraints. As an illustration, the four individuals (with $\ell = 2$ loci) $(\{1,2\}, \{1,1\})$, $(\{4,3\}, \{6,6\})$ and $(\{1,2\}, \{1,6\})$ form a full sibling group since they can be the children of the two parents $(\{1,3\}, \{1,6\})$ and $(\{2,4\}, \{1,6\})$.

Given these Mendelian constraints, our goal is to cover the universe $\mathcal{U}$ by a set of full-sibling groups under the parsimonious assumption of a minimum number of parents. Formally, the MIN-PARENT$_{n,\ell}$ problem is defined as follows.

**Problem name:** MIN-PARENT$_{n,\ell}$

**Input:** Our input is an universe $\mathcal{U}$ of $n$ individuals each with $\ell$ loci.

**Valid Solutions:** a cover $\mathcal{A}$ of $\mathcal{U}$ such that each set $S \in \mathcal{A}$ in the cover is a sibling group.

**Notation:** $\mathcal{B}(\mathcal{A})$ denote a set of individuals (parents) such that every set $S$ (sibling group) in the cover has its two parents from $\mathcal{B}(\mathcal{A})$.

**Objective for minimization:** $minimize\ |\mathcal{B}(\mathcal{U})| = \min_\mathcal{A} |\mathcal{B}(\mathcal{A})|$

In the setting of the implicit set cover problems described before, our cover problem is as follows:

- Our sets (sibling groups) are defined implicitly by the Mendelian constraints; note that the number of such sets is possibly exponential and thus we cannot always enumerate them in polynomial time.

- Our polynomial time oracle $\mathcal{O}$ answers queries of the following type: *given a given subset $\mathcal{U}' \subseteq \mathcal{U}$ of the universe, does $\mathcal{U}'$ form a valid (sibling) set following the Mendelian constraints*[2]? It is easy to show a polynomial-time implementation of the oracle (*e.g.*, see [4]).

Finally, note that our objective function is obviously monotone since $\mathcal{U}' \subset \mathcal{U}$ implies $|\mathcal{B}(\mathcal{U}')| \leq |\mathcal{B}(\mathcal{U})|$. *A natural parameter of interest in covering problems the maximum size (number of elements) $a$ in any set.* For our problem, the parameter $a$ corresponds to maximum number of individuals of any sibling group.

We first show that the MIN-PARENT problem is MAX-SNP-hard even if $a = 3$. This leads us to the question about the computational complexity of the problem for arbitrary $a$. We will show that, for arbitrary $a$, it is very hard to even find an approximation to a minimum set of parents for a *given* sibling partition of the universe with given a candidate set of parents that includes an optimal set of parents. Formally, the FIND-MIN-PARENT$_{n,\ell}$ is defined as follows.

**Problem name:** FIND-MIN-PARENT$_{n,\ell}$.

**Input:** a partition $\mathcal{A}$ of a set $\mathcal{U}$ of $n$ elements, each with $\ell$ loci, such that each set $S$ in the partition $\mathcal{A}$ is a sibling set, and a set of elements (possible parents) $\mathcal{P}$.

**Valid Solutions:** any $\mathcal{B}(\mathcal{A})$ provided that $\mathcal{B}(\mathcal{A}) \subseteq \mathcal{P}$.

**Notation:** $\mathcal{B}(\mathcal{A})$ denote a set of individuals (parents) such that every set $S$ (sibling group) in the cover has its two parents from $\mathcal{B}(\mathcal{A})$.

**Objective for minimization:** $minimize\ |\mathcal{B}(\mathcal{U})| = \min_{\mathcal{B}(\mathcal{A}) \subseteq \mathcal{P}} |\mathcal{B}(\mathcal{A})|$.

---

[2]Note that if $\mathcal{U}'$ is not a valid set, the oracle $\mathcal{O}$ does not provide any hint about other possible valid sets.

## 3.1 Standard Terminologies

Recall that a $(1+\varepsilon)$-*approximate solution* (or simply an $(1+\varepsilon)$-approximation) of a minimization (resp. maximization) problem is a solution with an objective value no larger (resp. no smaller) than $1+\varepsilon$ times (resp. $(1+\varepsilon)^{-1}$ times) the value of the optimum, and an algorithm achieving such a solution is said to have an *approximation ratio* of at most $1+\varepsilon$. A problem is $r$-inapproximable under a certain complexity-theoretic assumption means that the problem does not have a $r$-approximation unless the complexity-theoretic assumption is false.

L-reductions are a special kind of approximation-preserving reduction that can be used to show MAX-SNP-hardness of an optimization problem. Given two optimization problems $\Pi$ and $\Pi'$, $\Pi$ L-reduces to $\Pi'$ if there are three polynomial-time procedures $T_1, T_2, T_3$ and two constants $a$ and $b > 0$ such that the following two conditions are satisfied:

**(1)** For any instance $I$ of $\Pi$, algorithm $T_1$ produces an instance $I' = f(I)$ of $\Pi'$ generated from $T_1$ such that the optima of $I$ and $I'$, $OPT(I)$ and $OPT(I')$, respectively, satisfy $OPT(I') \leq a \cdot OPT(I)$.

**(2)** For any solution of $I'$ with cost $c'$, algorithm $T_2$ produces another solution with cost $c''$ that is no worse than $c'$, and algorithm $T_3$ produces a solution of $I$ of $\Pi$ with cost $c$ (possibly from the solution produced by $T_2$) satisfying $|c - OPT(I)| \leq b \cdot |c'' - OPT(I')|$.

An optimization problem is MAX-SNP-hard if another MAX-SNP-hard problem L-reduces to that problem. Arora et al. [1] show that, assuming P$\neq$NP, every MAX-SNP-hard problem is $(1+\varepsilon)$-inapproximable for some constant $\varepsilon > 0$ unless P=NP.

## 3.2 Our Results

For MIN-PARENT$_{n,\ell}$, we show in Section 4 that the problem is MAX-SNP-hard even if $a = 3$ and observe in Section 5 that for any $a$ and any integer constant $c > 0$ the problem admits an easy $\left(\frac{a}{c} + \ln c\right)\sqrt{n}$-approximation. We show in Section 6 that, for arbitrary $a$, FIND-MIN-PARENT$_{n,\ell}$ admits *no* $2^{\log^\varepsilon n}$-approximation, for *every* constant $0 < \varepsilon < 1$, unless NP$\subseteq$DTIME($n^{poly \log(n)}$).

# 4 Inapproximability of MIN-PARENT for $a = 3$

**Lemma 1** *MIN-PARENT$_{n,\ell}$ is MAX-SNP-hard even if $a = 3$.*

**Proof.** For notational simplification, when an individual has the multiset $\{x, x\}$ in a locus, we will refer to it by saying that the individual has a "**label**" of value $x$ in that locus. Our construction will ensure that all individuals have only one label at every locus. It is then easy to check that a set of individuals can be a sibling set if and only if at each locus they have labels with no more than two distinct values. In the sequel, we will use the terminologies "label $x$" and "locus $\{x, x\}$" interchangeably.

The (*vertex-disjoint*) triangle-packing (TP) problem is defined as follows. We are given an undirected connected graph $G$. A triangle is a cycle of 3 nodes. The goal is to find (pack) a maximum number of *node-disjoint* triangles in $G$. TP is known to be MAX-SNP-hard even if every vertex of $G$ has degree at most 4 [7]. Moreover, the proof in [7] show that the MAX-SNP-hard instances of TP in their reduction produces an instance of TP with $n$ nodes in which an optimal solution has $\alpha n$ triangles for some constant $0 < \alpha < 1$.

We will provide an approximation preserving reduction from an instance graph $G$ of $n$ nodes of TP with nodes of $G$ having a maximum degree of 4 as obtained in [7] to MIN-PARENT$_{n,\ell}$. We introduce an individual **u** for every node $u$ of the graph $G$ and provide ordered label sequences for each node (individual) such that:

**(1)** Three individuals corresponding to a triangle of $G$ have at most two values in every locus and thus can be a sibling set.

**(2)** Three individuals that do not correspond to a triangle of $G$ have at least three values in some locus and thus cannot be a sibling set.

**(3)** Consider any *maximal* set of vertex disjoint triangles in $G$ and the corresponding sibling sets (each of size 3). Partition the remaining vertices of $G$ not covered by these triangles arbitrarily into pairs (groups of size 2) and consider the corresponding full sibling sets (each of size 2). Then, each sibling set in the above collection requires two *new* parents.

Note that since we have a maximal set of triangles, no three vertices in the set of pairs can form a triangle. Conversely, given any solution of MIN-PARENT$_{n,\ell}$, we preprocess the solution to get a canonical solution to ensure that no three individuals in the union of pairs can be a sibling set; this preprocessing does not increase the number of sibling sets.

Note that, since any pair of individuals can be a full sibling set, the above properties imply that TP has a solution with $t$ triangles if and only if the MIN-PARENT problem can be solved with $2t + 2 \cdot \frac{n-3t}{2} = n - t$ parents.

The MAX-SNP-hardness now follows easily since an optimum solution of TP on $G$ has $\alpha n$ triangles for some constant $0 < \alpha < 1$. More precisely, let $I$ and $I'$ be the instance of TP and the corresponding instance of MIN-PARENT$_{n,\ell}$, respectively, and let OPT($I$) and OPT($I'$) denote the number of triangles and the number of parents in an optimal solution of $I$ and $I'$, respectively. Then, the following two statements hold.

**(a)** Since OPT($I$) = $\alpha n$ we have OPT($I'$) = $n - \alpha n = \frac{1-\alpha}{\alpha} \alpha n = \left(\frac{1-\alpha}{\alpha}\right)$ OPT($I$) where $\frac{1-\alpha}{\alpha}$ is a positive constant.

**(b)** Since OPT($I'$) = $n - \alpha n$ we must have $c' \geq n - \alpha n$. Thus, if $c' = n - \alpha n + x$ (for some $x$) is the number of parents in a solution of the instance $I'$ after preprocessing then number of triangles in the solution of the instance $I$ of TP is given by $c = n - c' = \alpha n - x$ and thus $|c - OPT(I)| = |c' - OPT(I')|$.

Now, we describe the reduction.

Our first set of loci are as follows. The index of a locus, which we call the "coordinate", is defined by an "origin" node $u$. Thus, we will have $|V|$ such loci, one for every node $u$. The respective label of an individual **v** at this coordinate is the *distance* (number of edges in a shortest path) from $u$ to $v$.

Our second set of loci are as follows. We have such a locus *for every* set of three vertices $\{u, v, w\}$ that *does not* form a triangle. Thus, we will have $O(|V|^3)$ such loci. Since the three vertices do not form a triangle, at least one pair of them, say $u$ and $v$, are not connected by an edge. As a result, the set of vertices $\{u, v, x\}$ do not form a triangle for any other vertex $x \notin \{u, v\}$. Our goal is to ensure that the vertices $u$, $v$ and $w$ cannot be a sibling group while not disallowing any other sibling groups that can be formed by a triangle in the graph. This is easy to do. Put the label 1 in this locus for the individual **u**, label 2 for individual **v** and label 3 for all other individuals.

First we need to check that Property **(1)** holds. The following is true with respect to the first set of loci. Consider a triangle $\{u, v, w\}$, any locus (coordinate) $\ell$ and assume that $u$ has the minimum label value of $L$, *i.e.*, it is *nearest* to the origin node that defined $\ell$. Then labels of $v$ and $w$ are at least $L$ and at most $L + 1$, hence $u$, $v$ and $w$ have at most two labels at $\ell$. The second set of loci never disallows a sibling group corresponding to a triangle, so the property is not violated by them either.

The construction of the second set of loci implies that Property **(2)** is true.

Finally, we need to verify Property **(3)**. There are three cases to verify.

First, consider the case when we have two sibling groups correspond to two triangles $T_1 = \{u, v, w\}$ and $T_2 = \{p, q, r\}$ in $G$. Note that since nodes in $G$ have a maximum degree of 4, any node of one triangle can be connected to at most two nodes in the other triangle.

The locus $\ell$ defined by the origin node $u$ has a label 0 for $u$ and a label 1 for $v$ and $w$. Thus, the sibling set $\{\mathbf{u}, \mathbf{v}, \mathbf{w}\}$ can be generated *only* by a pair of parents, say $A$ and $B$, each of which has the alleles $\{0, 1\}$ in locus $\ell$.

Since $u$ is connected to at most two nodes in $T_2$, it is not connected to a node in $T_2$, say $r$. Then, $r$ must have a label $x \geq 2$ in locus $\ell$. Thus, neither $A$ nor $B$ can be a parent of the sibling group $\{\mathbf{p}, \mathbf{q}, \mathbf{r}\}$ since $x \notin \{0, 1\}$.

Second, consider the case when the we have two sibling groups corresponding to a triangle $T = \{u, v, w\}$ and a pair $P = \{p, q\}$. Consider the locus defined by the origin node $u$. We have a label 0 for $u$ and a label 1 for $v$ and $w$ in this locus. Thus, the sibling set $\{\mathbf{u}, \mathbf{v}, \mathbf{w}\}$ can be generated *only* by a pair of parents, say $A$ and $B$, each of which has the alleles $\{0, 1\}$ in this locus. If node $u$ is *not* connected to both nodes $p$ and $q$ then one of the nodes which is not connected to $u$, say $p$, must have a label $x \geq 2$ in this locus. Thus, neither $A$ nor $B$ can be a parent of the sibling group $\{\mathbf{p}, \mathbf{q}\}$ since $x \notin \{0, 1\}$. Otherwise, it must be the case that $u$ is connected to *both* $p$ and $q$.

Repeating the same argument with $q$ as the origin node and then $r$ as the origin node shows that the only case that remains to be considered is when *each* of $u$, $v$ and $w$ is connected to *both* the nodes $p$ and $q$. But, then the induced subgraph of $G$ with vertices $u$, $v$, $w$, $p$ and $q$ is a 5-clique. Since every node in $G$ has a degree of no more than 4, this implies that $G$ has more than one connected component, contradicting the fact that $G$ was a connected graph.

Finally, consider the case when we have two sibling groups corresponding to two pairs $P_1 = \{u, v\}$ and $P_2 = \{p, q\}$. Since we have preprocessed the solution of MIN-PARENT$_{n,\ell}$ or equivalently have a maximal set of triangles for the solution of TP, node $u$ is not connected to at least one node in $P_2$, say $p$. The locus defined by the origin node $u$ has a label 0 for **u** and a label 1 for **v**, but has a label $x \geq 2$ for **p**. Thus, the sibling set $\{\mathbf{u}, \mathbf{v}\}$ can be generated *only* by a pair of parents, say $A$ and $B$, each of which has the alleles $\{0, 1\}$ in the corresponding locus, but neither $A$ nor $B$ can be a parent of the sibling group $\{\mathbf{p}, \mathbf{q}\}$ since $x \notin \{0, 1\}$. ❏

## 5 A Simple Approximation Algorithm for MIN-PARENT$_{n,\ell}$

Note that we do not need to know the value of $a$ in the theorem below.

**Observation 2** *Let $a$ be the maximum size of any sibling set. Then, for any integer constant $c > 0$, MIN-PARENT admits an easy $\left(\dfrac{a}{c} + \ln c\right)\sqrt{n}$-approximation with polynomially many access to the oracle $\mathcal{O}$ (and, thus in polynomial time).*

**Proof.** Our proof is similar to the analysis of a standard greedy algorithm for set cover problems [11].

Suppose that we have a subset $\mathcal{U}' \subset \mathcal{U}$ of the universe that is still not covered. We can enumerate all subsets of $\mathcal{U}'$ of size at most $c$ in $O(n^c)$ time and for each subset query the oracle $\mathcal{O}$ to find if any of these subsets of individuals are full siblings for the MIN-PARENT$_{n,\ell}$ problem. Thus we can assume that for every instance of the problem, either the maximum sibling set size is below $c$ and we can find such a group of maximum size, or we can find a sibling set of size $c$. Our algorithm simply selects such a set, removes the corresponding elements from $\mathcal{U}'$ and continues until all elements of $\mathcal{U}$ are covered.

Obviously, all subsets of a sibling set are valid sibling sets too. Let OPT be the minimum number of parents in an optimal solution of MIN-PARENT$_{n,\ell}$. Consider an optimum solution, make it disjoint by arbitrarily shrinking each full-sibling set and let $\alpha$ be the number of sets in this partition. Obviously, $\alpha \leq n/2$. Since no two full-sibling sets are produced by the same pair of parents (because of minimality), $\binom{\text{OPT}}{2} \geq \alpha$ which implies OPT$> \sqrt{2\alpha}$. We distribute the cost of our solution among the sets of the optimum. When a set with $b$ elements is selected, we remove each of its element and charge the sets of the optimum $1/b$ for each removal. It is easy to see that a set with $a$ elements will get the sequence of charges with values at most $(\underbrace{1/c, \ldots, 1/c}_{a-c \text{ times}}, 1/(c-1), 1/(c-2), \ldots, 1)$
and these charges add to $\frac{a}{c} - 1 + \sum_{i=1}^{c} \frac{1}{i} = \frac{a}{c} + \sum_{i=2}^{c} \frac{1}{i} < \frac{a}{c} + \ln c$. Thus, we use at most $\left(\frac{a}{c} + \ln c\right) \alpha$ sibling groups. Each sibling group can be generated by at most two new parents. Thus, the total number of parents necessary to generated these sibling groups is at most $\left(\frac{a}{c} + \ln c\right) \sqrt{2\alpha}$ OPT $< \left(\frac{a}{c} + \ln c\right) \sqrt{n}$ OPT. □

## 6 Inapproximability of FIND-MIN-PARENT

**Lemma 3** *For every constant $0 < \varepsilon < 1$, FIND-MIN-PARENT$_{n,\ell}$ admits no $2^{\log^\varepsilon n}$-approximation unless NP$\subseteq$DTIME$(n^{poly \log(n)})$.*

**Proof.** We first need the MINREP problem which is defined as follows. We are given a bipartite graph $G = (A, B, E)$. We are also given a partition of $A$ into $|A|/\alpha$ equal-size subsets $A_1, A_2, \ldots, A_\alpha$ and a partition of $B$ into $|B|/\beta$ equal-size subsets $B_1, B_2, \ldots, B_\beta$. These partitions define a natural "bipartite super-graph" $H$ in the following manner. $H$ has a "super-vertex" for every $A_i$ (the left partition) and a "super-vertex" for every $B_j$ (the right partition). There exists an "super-edge" between the super-vertex $A_i$ and the super-vertex $B_j$ if and only if there exists $u \in A_i$ and $v \in B_j$ such that $\{u, v\}$ is an edge of $G$. A pair of vertices $u$ and $v$ "witnesses" a super-edge $\{A_i, B_j\}$ provided $a \in A_i$, $b \in B_j$ and the edge $\{a, b\}$ exists in $G$. A set of vertices $S$ of $G$ witnesses a super-edge if there exists at least one pair of vertices in $S$ that witnesses the super-edge. The goal of the MINREP problem is to find $A' \subseteq A$ and $B' \subseteq B$ such that $A \cup B$ witnesses *every* super-edge of $H$ and the size of the solution, namely $|A'| + |B'|$, is *minimum*.

For notation simplicity, let $n = |A| + |B|$. The following result is a consequence of Raz's parallel repetition theorem [8, 9]. Let $L \in NP$ and $0 < \varepsilon < 1$ be any fixed constant. Then, there exists a reduction running in quasi-polynomial time, namely in time $n^{\text{poly} \log(n)}$, that given an instance $x$ of $L$ produces an instance of MINREP such that if $x \in L$ then MINREP has a solution of size at most at most $\alpha + \beta$, but if $x \notin L$ then MINREP has a solution of size at least $(\alpha + \beta) \cdot 2^{\log^\varepsilon n}$. Thus, the above theorem shows that MIN-REP has no $2^{\log^\varepsilon n}$-approximation under the complexity-theoretic assumption of NP$\not\subseteq$DTIME$(n^{\text{poly} \log(n)})$.

Let $L$ be any language in NP. Use the above theorem to translate an instance $x$ of $L$ to an instance of MINREP as described above. Now, we describe a translation of this instance of MINREP to an instance of FIND-MIN-PARENT$_{\mathcal{P}, n, \ell}$.

We have a parent $p_v$ in $\mathcal{P}$ corresponding to every element $v \in A \cup B$. We have an individual $s_{a,b}$ in $\mathcal{U}$ for every edge $\{a, b\}$ in $G$. Thus, the number of possible parents in $\mathcal{P}$ is $n$ and the number of individuals in $\mathcal{U}$ is $O(n^2)$. It therefore suffices to prove a $2^{\log^\varepsilon |\mathcal{P}|}$-inapproximability since that implies as $2^{\log^\varepsilon |\mathcal{U}|}$-inapproximability.

Before describing our reduction, we need a generic construction of the following nature to simplify our description. We are given two elements $p_u, p_v \in \mathcal{P}$ and an element $s_{a,b} \in \mathcal{U}$. We want to add a new locus with appropriate allele values to ensure that $s_{a,b}$ *cannot* be a child of $p_u$ and $p_v$, but *no other parent-child relationship is forbidden*. This is easy to do. Put the alleles $\{a, b\}$ in this locus for $p_u$ and $p_v$ and put the alleles $\{a, c\}$ in this locus for every individual (including $s_{a,b}$) in $(\mathcal{P} \cup \mathcal{U}) \setminus \{p_u, p_v\}$. It follows that $s_{a,b}$ cannot be a child of $p_u$ and $p_v$ since $c \notin \{a, b\}$, but no other child-parent combination is forbidden since $\{a, c\}$ can be produced by the Mendelian rule either from $\{a, b\}$ and $\{a, c\}$ or from $\{a, c\}$ and $\{a, c\}$.

Now, we add additional loci to the individuals in $\mathcal{U} \cup \mathcal{P}$ in the following manner following the two rules:

**Rule ($\star$):** For every edge $\{u, v\}$ of $G$ with $u \in A_i$ and $v \in B_j$ and for every pair of vertices $\{a, b\}$ such that $\{a, b\} \in E \setminus \{\{y, z\} \mid y \in A_i, z \in B_j, \{y, z\} \in E\}$ we add an additional locus using the generic construction to ensure that $s_{a,b}$ *cannot* be a child of $p_u$ and $p_v$.

**Rule ($\star\star$):** For every pair of vertices $u$ and $v$ of $G$ such that $\{u, v\} \notin E$ and for every pair of vertices $a$ and $b$ of $G$ such that $\{a, b\} \in E$, we add an additional locus using the generic construction to ensure that the individual $s_{a,b} \in \mathcal{U}$ *cannot* be a child of the parents $p_u$ and $p_v$ in $\mathcal{P}$.

We build each individual in $\mathcal{U} \cup \mathcal{P}$ locus-by-locus in the above manner. Our partition $\mathcal{A}$ of $\mathcal{U}$ to sibling groups is defined as follows: we have a sibling group $\mathcal{A}_{i,j} = \{\{s_{a,b}\} \mid \{a, b\} \text{ witnesses the super-edge } \{A_i, B_j\}\}$ for every super-edge $\{A_i, B_j\}$.

First, we need to verify that each of our sibling set is indeed a sibling set. Consider the sibling set $\mathcal{A}_{i,j}$. Pick any $u \in A_i$ and $v \in B_j$ such that $\{u, v\} \in E$, i.e., $\{u, v\}$ witnesses the super-edge $\{A_i, B_j\}$. We claim that $p_u$ and $p_v$ are the parents for all individuals in $\mathcal{A}_{i,j}$. Indeed, the two rules allow this.

Suppose that MINREP has a solution of size $\gamma$. This generates a set of $\gamma$ parents for FIND-MIN-PARENT in an obvious manner: for every vertex $v$ in the solution of MINREP we pick the individual $p_v$ in the solution of FIND-MIN-PARENT. If the super-edge $\{A_i, B_j\}$ is witnessed by the edge $\{u, v\}$ in the solution of MINREP, then the sibling set $\mathcal{A}_{i,j}$ is generated by the parents $p_u$ and $p_v$.

Conversely, suppose that FIND-MIN-PARENT has a solution with $\gamma$ parents. We associate each parent $p_u$ to the corresponding vertex $u$ of $G$ in our solution of MINREP. Consider a super-edge $\{A_i, B_j\}$ and the associated sibling set $\mathcal{A}_{i,j}$. Suppose that $p_u$ and $p_v$ are the parents of this group. By Rule ($\star\star$), $\{u, v\} \in E$. By Rule ($\star$), one of $p_u$ and $p_v$, say $p_u$, must be from $A_i$ and the other one $p_v$ from $B_j$. Thus, the edge $\{u, v\}$ witnesses this super-edge. ❑

**Remark 1** *The above reduction works even if one does not specify the set $\mathcal{A}$ of sibling partition explicitly as part of input but allows all feasible partitions.*

**Acknowledgements.** We thank Richard Karp for his talk at ICS-2009 that motivated us to think about our problem as an implicit cover problem.